\documentstyle[aps,pre,epsf]{revtex}

\newcommand{\be}{\begin{equation}}
\newcommand{\ee}{\end{equation}}
\newcommand{\bea}{\begin{eqnarray}}
\newcommand{\eea}{\end{eqnarray}}
\newcommand{\hh}{\tilde{h}}
\newcommand{\prt}{\partial}

\newcommand{\rgl}{\rangle}
\newcommand{\lgl}{\langle}
\newcommand{\clH}{{\cal H}}

\newcommand{\sinc}{\mbox{\rm sinc}\,}
\newcommand{\e}{\epsilon}
\newcommand{\enu}{\frac{\epsilon}{\nu}}
\begin{document}

\title{Semiclassical quantization of maps with a variable time scale}
\author{A. Iomin$^a$, S. Fishman$^a$, and G.M. Zaslavsky$^{b,c}$ \\
$^a$ Department of Physics, Technion, Haifa, 32000, Israel.  \\
$^b$ Courant Institute of Mathematical Sciences, \\
New York University, 251 Mercer Str., New York, NY 10012 \\
$^c$  Department of Physics, New York University, \\
2-4 Washington Place, New York, NY 10003. \\ }

\maketitle

\begin{abstract}
Quantization of energy balance equations, which describe  a
separatrix -- like motion is presented. The method is based on an exact
canonical transformation of the energy--time pair to the action-angle
canonical  pair, $ (E,t)\rightarrow (I,\theta) $.
Quantum mechanical dynamics can be studied in the framework of the new
Hamiltonian. This transformation also establishes a 
relation between a wide 
class of the energy balance equations and dynamical localization of classical
diffusion by quantum interference, that was studied in the field of quantum chaos. An  exact 
solution for a simple system is 
presented as well.

\noindent PACS numbers: 05.45.Mt, 03.65.Sq
\end{abstract}

\section{Introduction}
\def\theequation{1.\arabic{equation}}
\setcounter{equation}{0}
It is known for many applications \cite{zfil,zasl,LiLi,Ch79}
that classical dynamics near hyperbolic points can be described by
a map, $ \hat{\cal T} $ in the energy--time $ (E,t) $ canonical variables
$ \hat{\cal T}(E,t)\rightarrow (E,t) $.
This map is known as the separatrix map, and  defines a motion in the 
vicinity of a separatrix,
where a period of the unperturbed (twist) map is an arbitrary function
of energy $ T(E) $. Its explicit form is defined by $ n+2 $ turning points 
as follows \cite{zasl}
\be\label{feb1}
T(E)\sim\left\{ \begin{array}{ll}
\log\frac{1}{|E|} & \mbox{if $n=1$} \\
E^{-(n-1)/2}    & \mbox{if $n> 1$,} \end{array}
\right.
\ee
where the energy of the separatrix without loss of generality is taken
to be zero and it is assumed that $ |E|\ll 1$.
A perturbation that is a  periodic function of time, for 
instance,
$ \epsilon\sin\nu t $, that is relevant for a variety of applications, may be 
considered.
In this case, $ \hat{\cal T} $ is an energy balance equation that describes the
energy change over the period $ T(E) $:
\bea\label{f2}
E_{n+1}=E_n+\e\sin\nu t_n \nonumber \\
t_{n+1}=t_n+T(E_{n+1}),
\eea
where $ \nu $ is the frequency of the perturbation of the strength
$ \epsilon $, and it is assumed that $ E_n=E(t_n-0) $. The
period of the unperturbed nonlinear motion $ T(E_n) $  describes a wide 
class of nonlinear systems with variety of applications, including: the 
celestial mechanics of the perturbed Kepler system \cite{petrosky,sz87}, 
charge particles in a field of a wave packet \cite{zasl} leading to a 
separatrix mesh phenomenon in  non-KAM
(Kolmogorov-Arnold-Moser) systems \cite{zzs} ( with possible
applications for atom cooling traps \cite{gena}), and of electron
dynamics in superconducting Josephson junctions \cite{graham}.
Maps like (\ref{f2}) are also related  to a
description of Rydberg atoms in a microwave field by the Kepler map
\cite{dima,grah}. Also Bloch electrons in external fields are described by a
separatrix map \cite{geisel,iomin1,ifish,igf,iomin2} and similar systems.

The map (\ref{f2}) can be derived from the Hamiltonian,
\be\label{oct1}
H=H_0(E)+\tilde{V}(t,\tau)=H_0(E)+(\epsilon/\nu)\cos(\nu t)
\cdot\delta_{2\pi}(\tau),
\ee
where the unperturbed Hamiltonian
\be\label{feb2}
 H_0(E)=\frac{1}{2\pi}\int^ET(E')dE'
\ee
depends only on the energy $E$ that is conjugate to the time
$ t $, while $\tau$ is the formal time parameter, and
$ \delta_{2\pi}(\tau)\equiv\sum_{n=-\infty}^{\infty}
\delta(\tau-2\pi n) $ is the
periodic $\delta$-function with the period $ 2\pi $.
Quantization of the Hamiltonian (\ref{oct1}) in the framework of the
energy-time $(E,t)$ canonical pair has been presented in previous
studies (see for instance \cite{graham,dima,grah,heller}).
The main deficiency of quantization of the Hamiltonian (\ref{oct1}) is an
appearance of the unphysical time for a wave function and
this fact has been pointed out earlier, in publications \cite{dima,grah}.
Classically it is always possible to establish a link between
the formal time parameter $\tau$ and the real time $t$ for any individual
trajectory by the solution $ t=t(\tau) $ (or by canonical
transformation). In
quantum mechanics (for a wave function), it is no longer 
possible. Therefore, this approach of quantization in the framework of
$(E,t)$ variables was used to study time independent
characteristics, such as localization \cite{graham,dima,grah}. On the 
other hand, properties which are explicitly time dependent, such as 
dynamical correlation functions or
dynamical fluctuations, cannot be studied by this approach.
Moreover, there is no rigorous proof of the validity of these asymptotic 
solutions. It is also believed that the main deficiency  is related to the 
semiclassically approximate definition of the energy operator as
$\hat{E}=-i\hh\prt/\prt t $, where $\hh$ is the effective Planck's constant. 
There is no  physical justification of this semiclassical definition. 
Therefore, to overcome these shortcomings of the quantum description,
it is reasonable to rewrite the system in such a form, where the time
parameter appears as the physical time. It is convenient to rewrite the
system (\ref{f2}) and (\ref{oct1}) in 
terms of the action-angle variables $ (I,\theta)$, related by a canonical
transformation $ H(E,t,\tau) \rightarrow \clH(I,\theta,t) $, where
$ t $ is the real time. This canonical transformation must be
such that
under this variable change one  transforms $ H_0(E)\rightarrow\clH_0(I) $
and $ \tilde{V}(t,\tau)\rightarrow V(\theta,t) $.
It should be stressed that for kicked systems this change of variables has
some specific properties, namely, the new potential $ V(\theta,t) $
must be independent of $ I $ and it must be a function only of
the phase $ \theta $ and the time $t$. Otherwise a shift in action at a kick
is an ambiguous procedure due to the discontinuity of the action $
I(t=t_n) $  at the $n$-th kick.

In what follows, for the semiclassical quantization, the map (\ref{f2})
will be rewritten in the action--angle variables. In this case the
semiclassical quantization procedure is standard \cite{ifish,iomin2}.
Semiclassical quantization of area--preserving maps  was subject of 
earlier studies. In particular quantization of
monotonic twist maps \cite{usi}, with interpolating flows 
\cite{moser} was considered.
We show first, in Sec. 2, that the map (\ref{f2}) can be rewritten in terms of 
action--angle variables.
Semiclassical quantization and an exactly solvable problem are presented in
Sec. 3. The results are summarized  in Sec. 4.

\section{Exact Canonical Transformation $(E,t)\rightarrow (I,\theta)$ }
\def\theequation{2.\arabic{equation}}
\setcounter{equation}{0}
Hamiltonian equations of motion that produce the map (\ref{f2}) have the 
following formal form
\bea\label{for1}
&(a):~~~~~~~&dE/d\theta=-\prt H/\prt t=\e\sin\nu t
\sum_{n=-\infty}^{\infty}\delta(\theta-2\pi n) \nonumber \\
&(b):~~~~~~~&dt/d\theta=\prt H/\prt E=T(E)/2\pi,~~~~~~~~~~
\eea
where $H$ is given by (\ref{oct1}) and $\frac{d}{d\theta}$ is the formal 
time-derivative 
($\tau\equiv\theta$). Integration over $\theta$ in the limits $(2\pi n-0,
2\pi n +2\pi-0) $ gives (\ref{f2}).  We obtain from the 
equations (\ref{for1}a,b) the Hamiltonian equations for the action--angle
variables $ (I,\theta) $ and the real time parameter $ t $. Let us
start from (\ref{for1}b). Inverting this equation, one obtains
\be\label{for2}
\frac{d\theta}{dt}=\left[T(E)/2\pi\right]^{-1}=\Omega(I)\, ,
\ee
where it is supposed that $ T(E)\neq 0 $, that means that $ t $ is a 
single valued function of $ \theta $. The frequency 
$ \Omega(I) $ corresponds to some parametric variable change $ E=E(I) $
along a trajectory, such that 
\be\label{for3}
dE/dI=\Omega(I)\, .
\ee
In this case the energy $ E=\clH_0(I) $ is an unperturbed Hamiltonian with  
the 
new action variable $ I $, that is defined by (\ref{for3}). From 
(\ref{for1})--(\ref{for3}) we  perform 
the 
following chain of parametric changes 
\be\label{for4}
\frac{dE}{d\theta}=\frac{dt}{d\theta}\cdot\frac{dE}{dt}=\frac{dt}{d\theta}
\cdot\frac{dE}{dI}\cdot\frac{dI}{dt}=\dot{I}.
\ee
It follows from (\ref{for1}a) and (\ref{for4}) that
\be\label{for5}
\dot{I}=\e\sin\nu t\sum_{n=-\infty}^{\infty}\delta(\theta-2\pi n)=
\e\sin\nu t\frac{\prt}{\prt\theta}\Theta(\theta)\, ,
\ee
where $ \Theta(\theta) $ is a step function of the following form 
\be\label{for6}
\Theta(\theta)=\frac{\theta}{2\pi}+\frac{1}{\pi} 
\sum_{l=1}^{\infty}\frac{1}{l}\sin l\theta\, ,
\ee
and it is a result of integration of the periodic $\delta$-function
written in the form
\be\label{for7}
\sum_{n=-\infty}^{\infty}\delta(\theta-2\pi n)=
\frac{1}{2\pi}+\frac{1}{\pi}\sum_{l=1}^{\infty}\cos l\theta\, . 
\ee
Equations (\ref{for2}) and (\ref{for5}) are the Hamilton equations generated 
by the Hamiltonian
\be\label{for7a}
\clH=\clH_0(I)-2\e'\sin\nu t \left[\theta+
\sum_{l=1}^{\infty}\frac{1}{l}\sin l\theta \right].
\ee
where $\e'=\frac{\e}{2\pi}$. 

To restore periodicity of the Hamiltonian in $\theta$, the following 
gauge--like transformation is carried out
\be\label{for8}
J=I+\frac{\e'}{\nu} \cos\nu t\, .
\ee
The corresponding Hamiltonian is
\be\label{for10}
\clH=\clH_0(J- \frac{\e'}{\nu}\cos\nu t)-2\e'\sin\nu
t\sum_{l=1}^{\infty}\frac{1}{l}\sin l\theta\equiv\clH_0+V\, .
\ee
The equations of motion (\ref{for2}) and (\ref{for5}) take the  form 
\bea\label{for9}
&\dot{J}=2\e'\sin\nu t\sum_{l=1}^{\infty}\cos l\theta \nonumber \\
&\dot{\theta}=\Omega(J-\frac{\e'}{\nu}\cos\nu t)\, .
\eea
In this form, the equations were derived from the Hamiltonian $\clH$, 
therefore the transformation from 
$H$ 
to $\clH$, that was presented here, is a canonical transformation.

\section{Semiclassical quantization}
\def\theequation{3.\arabic{equation}}
\setcounter{equation}{0}
\subsection{Floquet operator}

The system (\ref{for9}),(\ref{for10}) can be simply quantized 
semiclassically with 
\[ J\rightarrow\hat{J}=-i\hh\prt/\prt\theta=\hh\hat{n},\]
where $\hh$, in our dimensionless units, is a dimensionless parameter, that 
plays the role of Planck's constant, while $\hat{n}$ is a quantum numbers
operator. 
In what follows $\hh$ will be used for the semiclassical quantization. 
Since the Hamiltonian (\ref{for10}) is periodic both in time and in 
the angle $\theta$, the Floquet theory can be used for the analysis.
Therefore, a solution of the Schr\"odinger equation
\be\label{f14}
i\hh\frac{\prt}{\prt t}\psi(\theta,t)=\hat{\clH}(\hat{n},\theta,t)
\psi(\theta,t)
\ee
can be considered due to the Floquet theorem in the following form
\be\label{f16}
\psi(\theta,t)=e^{-i\lambda t}
e^{i\kappa\theta}\psi_{\lambda,\kappa}(\theta,t)\, ,
\ee
where periodicity of $\hat{\clH}$ in both $\theta$ and $t$ is taken into 
account, and the functions
\be\label{pfun1} 
\psi_{\lambda,\kappa}\equiv\psi_{\lambda,\kappa}(\theta,t)=
\psi_{\lambda,\kappa} (\theta+2\pi,t+2\pi/\nu) 
\ee
are periodic in time with the periods of the 
perturbation $ 2\pi/\nu $ and periodic in $\theta$ with the period $2\pi$,
while $0\leq\kappa<1$ and $\lambda$ are a ``quasimomentum'' and a 
quasienergy correspondingly.
Taking into account that commutation of the Hamiltonian with the 
exponential $e^{i\kappa\theta}$ leads to the shift on $\kappa$ for
$\hat{n}$,
\be\label{commut}
 \hat{\clH}(\hat{n},\theta,t) 
e^{i\kappa\theta}\psi_{\lambda,\kappa}(\theta,t)=
e^{i\kappa\theta}\hat{\clH}(\hat{n}+\kappa,\theta,t)
\psi_{\lambda,\kappa}(\theta,t), 
\ee
we can rewrite the Schr\"odinger equation (\ref{f14})
in the following form 
\be\label{f15}
\hat{F}(\kappa)\psi_{\lambda,\kappa}=\hh\lambda\psi_{\lambda,\kappa},
\ee
where $F\equiv F(\kappa)$ is the so--called Floquet operator 
\be\label{f17}
\hat{F}(\kappa)=-i\hh\frac{\prt}{\prt 
t}+\hat{\clH}(\hat{n}+\kappa,\theta,t),
\ee
with $ \lambda $ and $ \psi_{\lambda,\kappa} $ as the eigenvalues and
eigenfunctions correspondingly. 
The solution is cast in the form of the Fourier expansion
\be\label{f20}
|\psi_{\lambda,\kappa}\rangle=\sum_{n,j}\phi_{n,j}|n\rangle|j\rangle\, ,
\ee
moreover the Fourier harmonics \be\label{f18}
|n,j\rangle\equiv|n\rangle|j\rangle =
\frac{\sqrt{\nu}}{2\pi}e^{in\theta}e^{-ij\nu t}
\ee
are the eigenfunctions of the unperturbed system with $\e=0$, and 
$\phi_{n,j}=\langle j|\langle n|\psi_{\lambda,\kappa}\rangle$ are the
coefficients of the expansion. Matrix elements
of the Floquet operator 
\be\label{for11}
F_{n,n'}^{j,j'}=\lgl j|\lgl n|\hat{F}|n'\rgl|j'\rgl
\ee
specify the equations for the coefficients $ \phi_{n,j} $. 
First we calculate the matrix elements for $\hat{\clH}_0$. For this purpose we 
rewrite it formally as
\be\label{for12}
\hat{\clH}_0(\hat{J}+\kappa-\frac{\e'}{\nu}\cos\nu t)=
e^{i\frac{\e'}{\nu}\cos\nu 
t(i\prt_{\kappa})}\hat{\clH}_0(\hat{J}+\kappa)\, ,
\ee
where $\prt_{\kappa}\equiv\prt/\prt\kappa$ and it is assumed that 
${\clH_0}$ does not operate on 
functions of $\kappa$.
Calculation of the matrix elements (\ref{for11}) yields for (\ref{for12})
\be\label{for13}
{\clH_0}_{n,n'}^{j,j'}=\lgl j|e^{i\frac{\e'}{\nu}
\cos\nu t(i\prt_{\kappa})}|j'\rgl
\lgl n|\hat{\clH}_0(\hat{J}+\kappa)|n'\rgl=
\sum_mi^mJ_m(i\frac{\e'}{\nu}\prt_{\kappa})\delta_{j',j+m}
\clH_0(n'+\kappa)\delta_{n',n}\, ,
\ee
where $ J_m(x) $ is the Bessel function.
Correspondingly, the matrix elements for the additive part of the 
perturbation in (\ref{for10}) are 
\be\label{for14}
V_{n,n'}^{j,j'}=\left\{ \begin{array}{r@{\quad:\quad}l}
\frac{1}{2}\frac{\e'}{n-n'}\left[\delta_{j',j+1}-\delta_{j',j-1}\right] 
& n\neq n' \\
0 & n=n' \end{array}\right.  \, .
\ee
These matrix elements of (\ref{for13}) and (\ref{for14}) lead to the 
following equation for the expansion coefficients
\be\label{for15}
\sum_mi^mJ_m(i\frac{\e'}{\nu}\prt_{\kappa})\clH_0(n+\kappa)\phi_{n,j+m}+
\frac{\e'}{2}\sum_{n' \neq n}\frac{1}{n-n'}[\phi_{n',j+1}-\phi_{n',j-1}]=
\hh(\lambda+\nu j)\phi_{n,j}\, .
\ee
It is similar to a quasi-1D Anderson-like chain for the
cases when  $ \clH_0(n+\kappa) $ corresponds to a random potential.
It is also the generalization of a specific application  in \cite{naama}.
The classical motion is chaotic in many types of $ T(E) $ in 
(\ref{f2}). Therefore, the obtained equation establishes a relation
between a wide class of energy balance equations of the form (\ref{f2}),
(\ref{oct1}) and dynamical localization of the classical chaotic diffusion 
\cite{FGP}. 
In the general case this equation is useful for numerical studies. It is
analytically tractable in perturbation theory for the parameter $\enu\ll 1$. 
Therefore, it is instructive to present an exactly solvable 
example where (\ref{feb2}) is linear, with a constant period $ 
T(E)=2\pi/\omega $. In this case a solution can be obtained in an explicit form 
analytically. 

\subsection{An exact model}

An exact solution can be obtained for the harmonic oscillator
$ \clH_0(I)=\omega I $ with constant frequency $ \Omega(I)=\omega $.
The linear driven system described
by the Hamiltonian $ \clH(I,\theta,t) $ is integrable.
The equation  (\ref{for15}) takes  the following simple form 
\be\label{for16}
\hh[\omega n-\nu j-\lambda+\omega \kappa]\phi_{n,j}-
\frac{\epsilon'\omega}{2\nu}[\phi_{n,j+1}+\phi_{n,j-1}]
+\frac{\e'}{2}\sum_{n' \neq n}\frac{1}{n-n'}[\phi_{n',j+1}-\phi_{n',j-1}]=0.
\ee
For the specific case of the harmonic oscillator $\kappa=0$. For other 
linear models, for example models of an  appropriate band structure in solids, 
$\kappa$
does not have to vanish.
A   solution of equation 
(\ref{for16}) is cast
in the form of a $ \sinc $ function by the following substitution
\be\label{for17}
\phi_{n,j}=Z_j\sinc[\pi(n-\lambda/\omega+\kappa)],
\ee
where the $ \sinc $ function is $ \sinc x=\frac{\sin x}{x} $ and $Z_j$ has 
to be determined.
In this case the quasienergy spectrum turns out to be
\be\label{for18}
\lambda=\omega \kappa+\omega n\, .
\ee 
Inserting (\ref{f17}) in (\ref{for16}) and taking into account
that 
\be\label{for19}
\sum_{n' \neq n}\frac{1}{n-n'}\sinc\pi(n'-n)=0,
\ee
since $ \sin\pi(n'-n)=0 $,
one obtains from (\ref{for16}) the following relation for the $Z_j$
\be\label{for20}
2jZ_j+z[Z_{j+1}+Z_{j-1}]=0,
\ee
where $ z=\epsilon' \omega /\hh\nu^2 $. 
The solution is \cite{yel}
\be\label{for21}
Z_j\equiv Z_j(z)=(-1)^jJ_j(z).
\ee

The result (\ref{for18}) and (\ref{for21}) is only a specific solution. 
The general solution is found by replacing 
in  (\ref{for16}) $j$ by $j-j_0$ and $\lambda$ by $\lambda-\nu j_0$. 
The resulting solution of the eigenvalue 
problem is 
\be\label{for22}
\lambda_{j_0}=\nu j_0+\omega \kappa+\omega n\, 
\ee
and the corresponding eigenfunction is
\be\label{for23}
Z_j=(-1)^{j-j_0}J_{j-j_0}(z).
\ee
Floquet theory implies that only $\lambda_{j_0}$ in an interval of size 
$\nu$, say $[0,\nu)$ should be used (see 
(\ref{f16})). In particular for the harmonic oscillator 
\be\label{for24}
\lambda_{j_0}=\nu j_0+\omega n\, ,
\ee
if it is in the interval $[0,\nu)$, as expected.

\section{Summary}

Quantization of energy balance equations, in  the
form of the map (\ref{f2}), is presented. This procedure of quantization
consists of the two steps. The first one is a transformation of the
Hamiltonian (\ref{oct1}) to the form (\ref{for10}) by the canonical
transformation $ (E,t)\rightarrow (I,\theta) $.
The second stage is semiclassical quantization.
The Hamiltonian formulation of the problem allows semiclassical
quantization of action--angle variables  in the framework of the standard 
procedure:
$I\rightarrow \hat{I}=-i\hh\prt/\prt\theta\equiv\hh\hat{n}$,
where $ \hh $ is a dimensionless semiclassical parameter such that
the classical limit is $ \hh\rightarrow 0 $.
It should be noted that the semiclassical approximation requires
that the width of the potential $V$ is larger than the
de-Broglie wavelength. Therefore, it is necessary to truncate
the Fourier expansion in (\ref{for6}), corresponding to replacement 
of the $\delta$-function by a function of finite width. 
The results obtained in Sec. 3.2 in the semiclassical approximation are 
correct only  up to the leading correction 
in 
$\hh$ to the classical ones \cite{ifish}.
The advantage
of the present analysis is that the transition from (\ref{f2}) to
(\ref{for10}) is exact, and wave functions depend on physical time,
while the main deficiency of straightforward quantization of (\ref{f2}) is
an appearance of an unphysical time parameter \cite{dima,grah}.

\section*{Acknowledgments}
We thank V. Afraimovich, I. Guarneri, A. Neishtadt and V. Rom-Kedar 
for fruitful discussions.
We also thank U. Smilansky for drawing our attention to the results
presented in \cite{usi,moser}.
This research is supported by U.S. Department of Navy Grant No
N00014-02-1-0056, the US--Israel Binational Science Foundation
(BSF) and by the Minerva Center of Nonlinear Physics of Complex Systems.

\end{document}